\documentclass[%
aps,prb,twocolumn,reprint,amsmath,amssymb,showpacs,superscriptaddress
]{revtex4-2}

\usepackage{epsfig}
\usepackage{amsmath}
\usepackage{graphics}
\usepackage{bm}
\usepackage{makecell}
\usepackage{float}
\usepackage{multirow}
\usepackage{dcolumn}
\usepackage{hyperref}
\hypersetup{
	colorlinks=true,
	linkcolor=blue,
	filecolor=gray,
	urlcolor=blue,
	citecolor=blue,
}
\begin{document}

\preprint{APS/123-QED}

\title{Superconductivity of Light-Elements Doped H$ {}_{3} $S}

\author{Hongyi Guan}
\author{Ying Sun}
\affiliation{International Center for Computational Method $\&$ Software and State Key Laboratory of Superhard materials, College of Physics, Jilin University, Changchun 130012, China 
}
\author{Hanyu Liu}
\email{hanyuliu@jlu.edu.cn}
\affiliation{International Center for Computational Method $\&$ Software and State Key Laboratory of Superhard materials, College of Physics, Jilin University, Changchun 130012, China 
}
\affiliation{Key Laboratory of Physics and Technology for Advanced Batteries (Ministry of Education), College of Physics, Jilin University, Changchun 130012, China}
\affiliation{International Center of Future Science, Jilin University, Changchun 130012, China}


\date{\today}

\begin{abstract}
Pressurized hydrogen-rich compounds, which could be viewed as precompressed metallic hydrogen, exhibit high superconductivity, thereby providing a viable route toward the discovery of high-temperature superconductors. Of particular interest is to search for high-temperature superconductors with low stable pressure in terms of pressure-stabilized hydrides. In this work, with the aim of obtaining high-temperature superconducting compounds at low pressure, we attempt to study the doping effects for high-temperature superconductive $ \mathrm{H_3S} $ with supercells up to 64 atoms using first principle electronic structure simulations. As a result of various doping, we found that Na doping for $ \mathrm{H_3S} $ could lower the dynamically stable pressure by 40 GPa. The results also indicate P doping could enhance the superconductivity of $ \mathrm{H_3S} $ system, which is in agreement with previous calculations. Moreover, our work proposed an approach that could reasonably estimate the superconducting critical temperature ($ T_{c} $) of a compound containing a large number of atoms, saving the computational cost significantly for large-scale elements-doping superconductivity simulations.
\end{abstract}

\maketitle


\section{INTRODUCTION}

The search for the high-temperature superconducting hydrides at high pressures has attracted attention in condensed matter physics field. In this regard, many hydrides with relatively high $ T_c $ were identified under high pressure \cite{doi:10.1063/1.4874158,RN4,RN1,RN9,Liu6990,PhysRevLett.119.107001,PhysRevLett.122.027001,RN8,Semenok_2020,kong2019superconductivity,PhysRevLett.123.097001,RN5,FLORESLIVAS20201,doi:10.1063/5.0033232,doi:10.1063/1.5079225}. Among these high-temperature superconducting hydrides, the theoretically predicted $ \mathrm{H_{3}S} $ with $ T_{c} $ of 203 K \cite{doi:10.1063/1.4874158,RN4,RN1} and $ \mathrm{LaH_{10}} $ with $ T_{c} $ of 250-260 K in \cite{Liu6990,PhysRevLett.119.107001,RN8,PhysRevLett.122.027001}, as well as CaH$_6$ being the first example of clathrate hydrides ever predicted \cite{Wang6463,ma2021experimental,li2021superconductivity}, were  experimentally synthesized. Recently, the carbonaceous sulfur hydride was found to possess extremely high superconductivity with a $ T_c $ as high as 288 K at 267 GPa \cite{RN5}, however, the actual crystal structure and the mechanisms of such extremely high superconductivity remains unclear and missing \cite{PhysRevB.101.174102,PhysRevB.101.134504,hu2020carbondoped,DOGAN20211353851,wang2021absence}.

Structure searches below 200 GPa were performed for the C-S-H system \cite{hu2020carbondoped,PhysRevB.101.174102,PhysRevB.101.134504,PhysRevB.103.L140105,du2021superconductivity}, while no high superconductive structure is yet identified. Later, it was reported that the high superconductivity in the C-S-H compound could be explained by the doping of C into the $ Im\bar{3}m $ $ \mathrm{H_3S} $ \cite{GE2020100330,hu2020carbondoped}, whereas these simulations are based on the virtual crystal approximation (VCA) \cite{https://doi.org/10.1002/andp.19314010507}. This approximation is employed with linearly mixed pseudopotentials and couldnot take the details of symmetry breaking and local distortions into account. These issues, in principle, could be addressed by performing the electronic simulations of doping carbon into a sufficiently large supercell of $ \mathrm{H_3S} $, if the computational power is allowed.   

In this paper, we investigated the structures and physical properties of partially substituting sulfur by the light elements in $ Im\bar{3}m $ $ \mathrm{H_3S} $ with supercell approach at the pressure range of 150-250 GPa by first-principle calculations. Our calculations mainly focus on $ \mathrm{H_{24}S_7X} $ and $ \mathrm{H_{48}S_{15}X} $, where $ \mathrm{X} $ denotes the doping elements from H to Cl without He and Ne in the periodic table. As a result, we found that the Na doing could lower the dynamically stable pressure of 40 GPa compared to the parent $ \mathrm{H_3S} $ system. Furthermore, the P doping could enhance the superconductivity of the parent $ \mathrm{H_3S} $ system, which is ascribed to octahedra units [SH${}_{6} $] and [PH${}_{6} $]. In addition, we proposed an estimation approach to investigate the low proportion C-doping effects at 260 GPa. The results suggest the estimated $ T_{c} $ is much lower than the room temperature.  


\section{COMPUTATIONAL DETAILS}

The structural optimization was done by the Vienna Ab initio Simulation Package (VASP) \cite{PhysRevB.54.11169}, with pseudopotentials employing generalized gradient approximation (GGA) based Perdew-Burke-Ernzerhof (PBE) type exchange correlation functional \cite{PhysRevLett.77.3865} and projector-augmented wave method \cite{PhysRevB.50.17953}. Monkhorst $ \boldsymbol{k} $ meshes \cite{PhysRevB.13.5188} spacing $ 2\pi \times 0.1 $ \r{A}${}^{-1}$ was used to sample the first Brillouin Zones. The electronic density of states was also computed by VASP with $ 20\times 20\times 20 $ $ \boldsymbol{k} $ mesh and was analyzed by VASPKIT \cite{wang2021vaspkit}. The phonon properties and superconducting properties were computed by the Quantum-Espresso (QE) package, with vanderbilt ultra-soft pseudopotentials \cite{PhysRevB.41.7892}. We have adopted $ \boldsymbol{k} $ mesh of $ 16\times 16\times 16 $ and $ \boldsymbol{q} $ mesh of $ 4\times 4\times 4 $ and tested the convergence with $ \boldsymbol{k} $ mesh of $ 20\times 20\times 20 $ and $ \boldsymbol{q} $ mesh of $ 5\times 5\times 5 $ for $ \mathrm{H_{24}S_7X} $. For $ \mathrm{H_{48}S_{15}X} $, $ \boldsymbol{k} $ mesh was used as $ 12\times 12\times 12 $ and $ \boldsymbol{q} $ mesh was used as $ 3\times 3\times 3 $. The smearing method was Methfessel-Paxton first-order spreading \cite{PhysRevB.40.3616} of 0.03 Ry. The cutoff energy for basis of plane waves was employed to be 100 Ry. Then, the transition temperatures were estimated by McMillan-Allen-Dynes (MAD) formula \cite{PhysRevB.12.905}
\begin{equation}
	T_c = f_{1}f_{2}\dfrac{\omega_{\mathrm{log}}}{1.20}\mathrm{exp}\left[-\dfrac{1.04(1+\lambda)}{\lambda-\mu^{*}(1+0.62\lambda)}\right] \label{mad-1}
\end{equation}
where 
\begin{equation}
	\begin{aligned}
		f_{1} &= \left\{1+\left[\dfrac{\lambda}{2.46(1+3.8\mu^{*})}\right]^{3/2}\right\}^{1/3}\\
		f_{2} &= 1+\dfrac{\lambda^2(\omega_2/\omega_{\text{log}}-1)}{\lambda^2+[1.82(1+6.3\mu^{*})(\omega_2/\omega_{\text{log}})]^2}
	\end{aligned} \label{mad-2}
\end{equation}
are the correction factors. $ \mu^{*} $, $ \lambda $ and $ \omega_{\text{log}} $ indicate the screened Coulomb parameter, electron-phonon coupling constant and the logarithm average over phonon frequency, respectively.

We have also computed the results by Migdal-Eliashberg (ME) theory \cite{migdal1958interaction,eliashberg1960interactions,PhysRev.148.263,PhysRevB.85.184514} 
\begin{gather}
	Z(i\omega_j) = 1+\dfrac{\pi T}{\omega_j}\sum\limits_{j'}\dfrac{\omega_{j'}}{\sqrt{\omega_{j'}^2+\Delta^2(i\omega_{j'})}}\lambda (i\omega_j-i\omega_{j'}) \\
	\begin{aligned}
		Z(i\omega_j)\Delta (i\omega_j) &=\pi T \sum\limits_{j'}\dfrac{\Delta (i\omega_{j'})}{\sqrt{\omega_{j'}^2+\Delta^2(i\omega_{j'})}} \\
		&  \times [\lambda (i\omega_j-i\omega_{j'})-\mu^{*}] 
	\end{aligned}  \\
	\lambda (i\omega_j-i\omega_{j'}) = \int \mathrm{d}\omega \dfrac{2\omega\alpha^2F(\omega)}{\omega^2+(\omega_j-\omega_{j'})^2}
\end{gather}
to compare with that of MAD equation, which is realized by the Elk code \cite{elk}. $ T_c $ could be obtained once the superconducting gap $ \Delta (i\omega_j) $ becomes zero in numerically solving ME equation.

\begin{figure}[!t]
	\centering
	\includegraphics[width=0.95\linewidth]{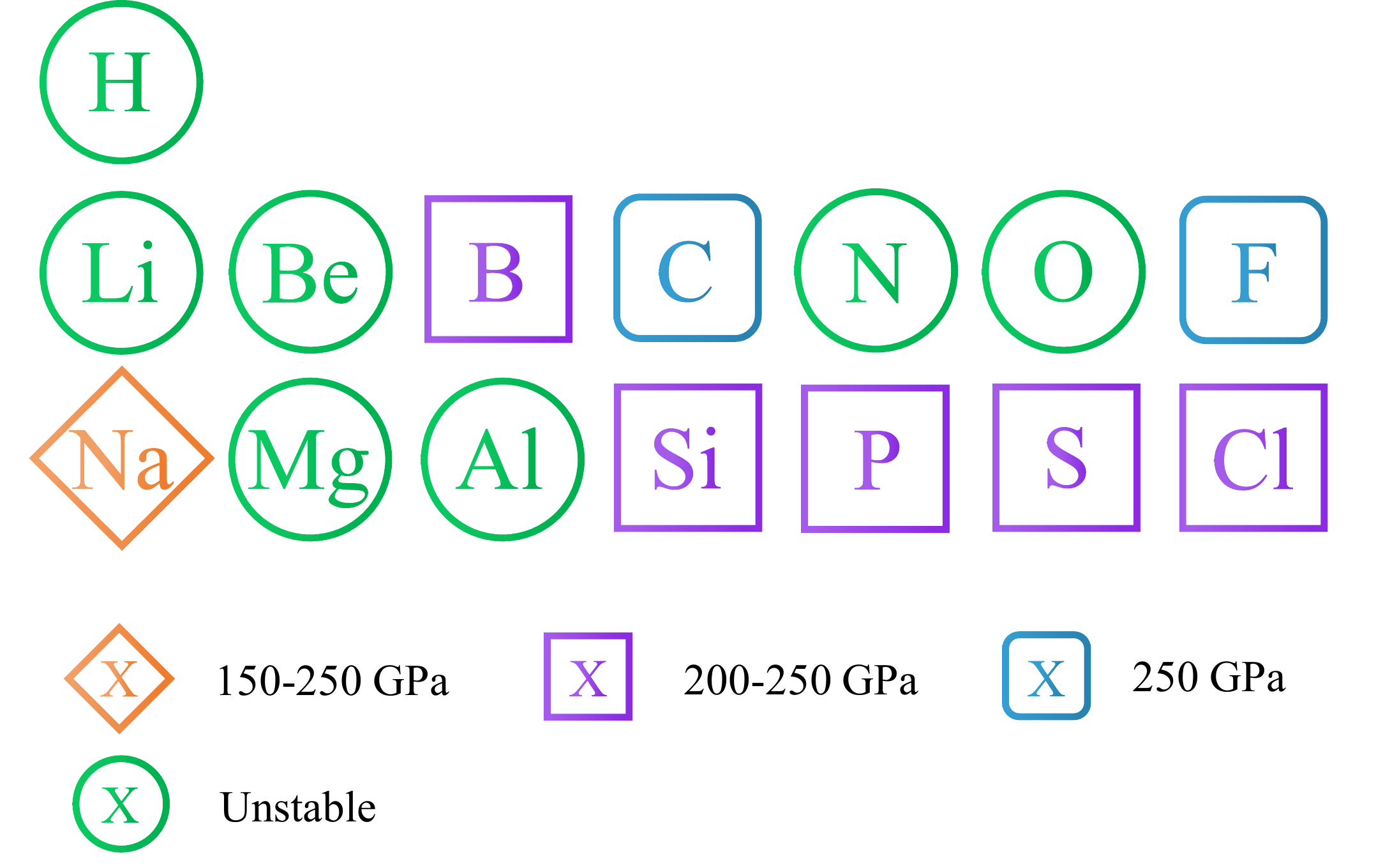}
	\caption{Summary of dynamical stability and stable pressure ranges of $ \mathrm{H_{24}S_7X} $, where X is the dopant.}
	\label{stability}
\end{figure}


\section{RESULTS AND DISCUSSIONS}
 
We began our simulations on investigating the validity of the supercell approach by computing the electronic properties and phonon properties of primitive cell and a supercell of 32 atoms ($ \mathrm{H_{24}S_{8}} $) for $ \mathrm{H_{3}S} $ at 200 GPa, as shown in Fig. S1 \cite{SI}. As shown in Fig. S2 \cite{SI}, the superconductivity using the supercell is well consistent with that simulated from primitive cell of $\mathrm{H_{3}S}$, which is also in agreement with previous results \cite{RN4,RN1}. The relevant information is listed in Table S1 \cite{SI}. 

\begin{figure}[!t]
	\centering
	\includegraphics[width=\linewidth]{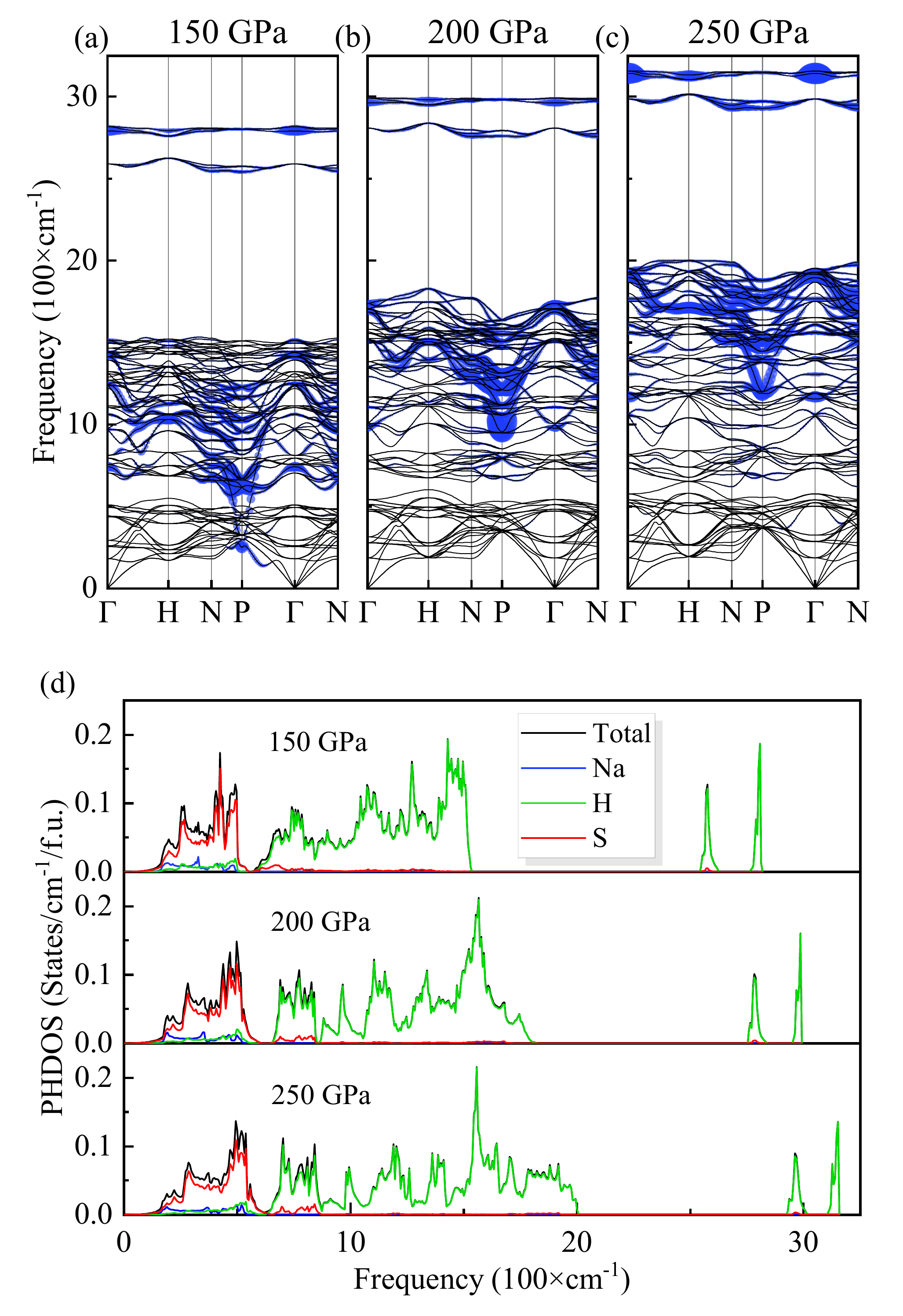}
	\caption{Phonon dispersion and phonon linewidth for $ \mathrm{H_{24}S_7Na} $ at pressure (a) 150 GPa, (b) 200 GPa and (c) 250 GPa, the magnitude of the phonon linewidth is indicated by the radii of blue circles. (d) Phonon density of states for $ \mathrm{H_{24}S_7Na} $ at 150-250 GPa.}
	\label{na_ph}
\end{figure}

\begin{figure*}[!ht]
	\includegraphics[width=\linewidth]{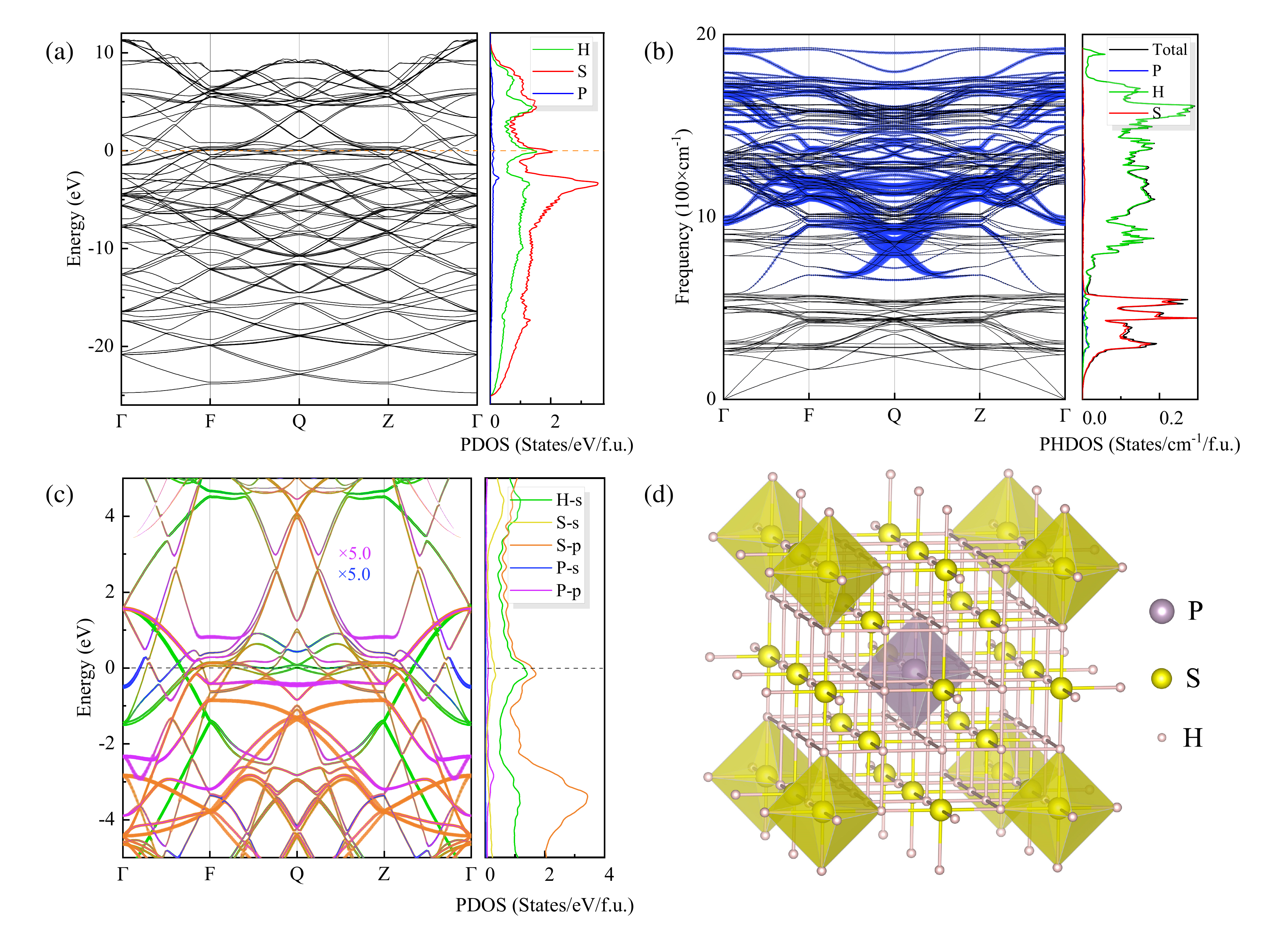}	 
	\caption{(a) Electronic band structure (left panel) and projected density of states (right panel) of $ \mathrm{H_{48}S_{15}P} $ at 200 GPa. (b) Phonon dispersion with phonon linewidth (left panel) and phonon density of states (right panel) of $\mathrm{H_{48}S_{15}P} $ at 200 GPa. The radii of the blue circles indicate the magnitude of the phonon linewidth. (c) Atom projected and orbital projected band structures (left panel) and density of states (right panel) of $\mathrm{H_{48}S_{15}P} $ at 200 GPa near the Fermi surface. The width of the lines indicates the weights of the corresponding orbitals. Due to the low proportion of P atoms, their weights are displayed fivefold. (d) Crystal structure of $\mathrm{H_{48}S_{15}P} $ at 200 GPa. The structures of [$ \mathrm{PH_{6}} $] and [$ \mathrm{SH_{6}} $] units are represented by the purple and yellow octahedra units respectively.}
	\label{p_cub}
\end{figure*}
Furthermore, we have systematically investigated the doping of $ \mathrm{H_{3}S} $ by the elements from H to Cl without He and Ne in the periodic table using supercells of 32 and 64 atoms. The structures of $ \mathrm{H_{24}S_7X} $ and $ \mathrm{H_{48}S_{15}X} $ are provided in Fig. S3 \cite{SI}. The results indicate the doping could destabilize the structure for several compounds. For the $ \mathrm{H_{24}S_7X} $ compounds, for example, imaginary frequency was found for $ \Gamma $ point with X=H, indicating dynamical instability. The dynamical stability of $ \mathrm{H_{24}S_7X} $ compounds within the range of 150-250 GPa is summarized in Fig. \ref{stability}. Moreover, we found that 12.5\% doping of Na into $ \mathrm{H_{3}S} $ has a lower dynamical stable pressure (140 GPa) compared to 180 GPa of the parent $ \mathrm{H_{3}S} $ \cite{RN4}. The absence of imaginary phonon frequency in the simulated phonon dispersion implies the dynamical stability of $ \mathrm{H_{24}S_7Na} $, as shown in Fig. \ref{na_ph}. It is clearly seen that the strongest electron-phonon interaction mainly emerges around $P$ point, which may lead to a large $ \lambda $ of 2.31. As a result, $ T_c $ of $ \mathrm{H_{24}S_7Na} $ can reach 191 K at 150 GPa by using ME equation with $ \mu^{*}=0.10 $.  
\begin{figure}[!hb]
	\centering
	\includegraphics[width=\linewidth]{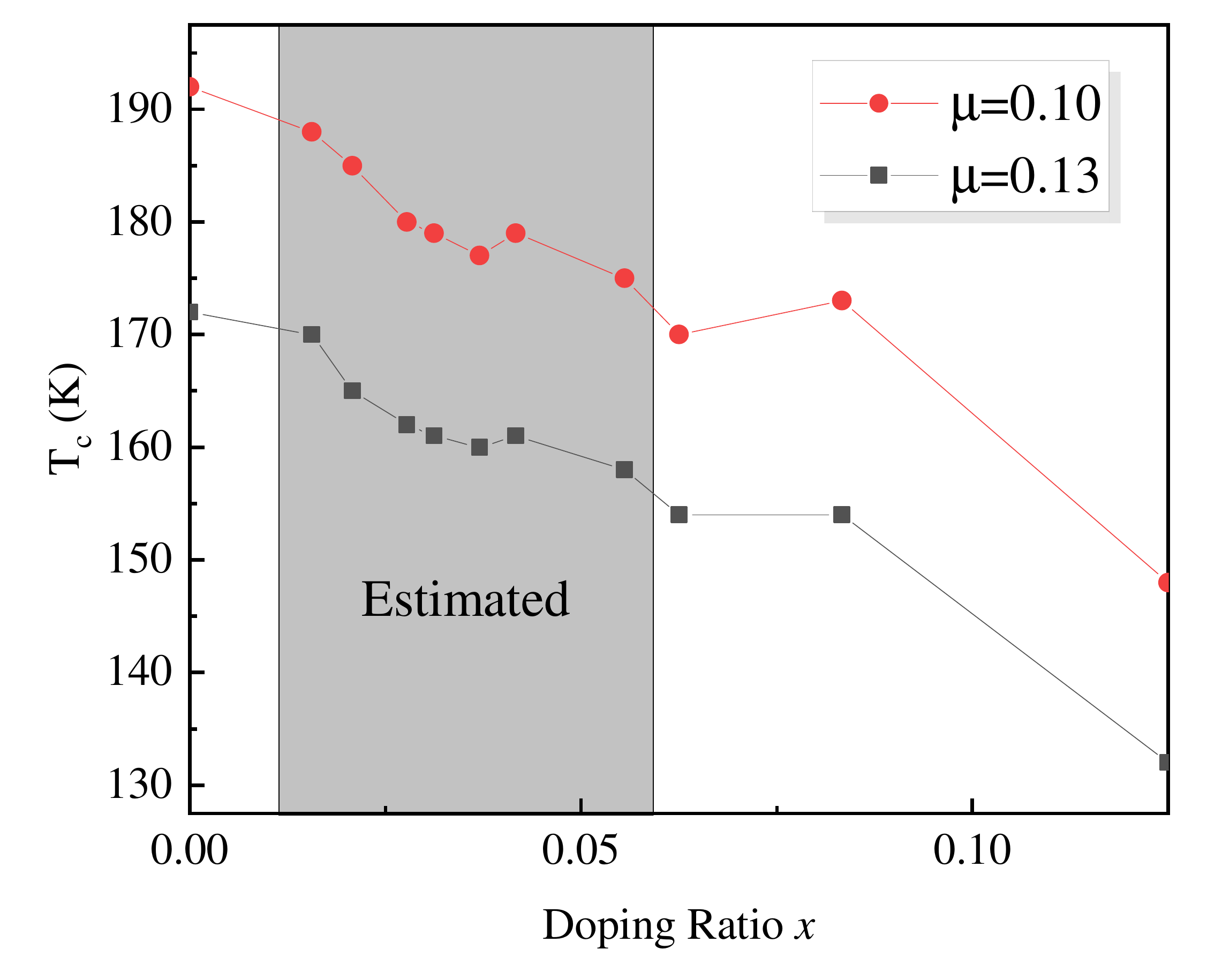}
	\caption{Calculated $ x $ dependence of $ T_{c} $ for H$ {}_{3} $S$ {}_{1-x} $C$ {}_{x} $ at 260 GPa. The points in the gray region are estimated by Eqs. (\ref{estimate1}), (\ref{estimate2}).}
	\label{c260}
\end{figure}

\begin{table*}[!ht]
	\centering
	\caption{The parameters of the superconductivity for the doped $ \mathrm{H_3S} $ at a pressure range of 150-250 GPa, where the typical screened Coulomb parameter $ \mu^{*} $ of 0.1-0.13 is employed.}
	\begin{ruledtabular}
		\begin{tabular}{cccccccc}
			Dopant & Doping ratio& Pressure (GPa) & $\lambda$ & $\omega_{\mathrm{log}}$ (K) & \makecell[c]{$N(0)$ \\ (States/Ry/atom)} & \makecell[c]{$T_c$ by \\ MAD equation (K)} & \makecell[c]{$T_c$ by \\ ME equation (K)} \\ \hline
			B & 0.1250 & 200 & 1.56  & 1281 & 0.721  & 154-171 & 176-191 \\
			B & 0.1250 & 250 & 1.21  & 1447 & 0.692  & 126-143 & 140-156 \\
			C & 0.1250 & 250 & 1.40  & 1259 & 0.709 & 134-150 & 153-169 \\
			F & 0.1250 & 250 & 1.04  & 1428 & 0.605  & 97-113 & 105-120 \\
			Na & 0.1250 & 150 & 2.31  & 846 & 0.656  & 154-169 & 178-191 \\  
			Na & 0.1250 & 200 & 1.56  & 1094 & 0.660  & 133-148 & 149-167 \\
			Na & 0.1250 & 250 & 1.42  & 1152 & 0.667  & 125-140 & 138-152 \\       
			Si & 0.1250 & 200 & 1.99  & 1155 & 0.838  & 182-199 & 210-225 \\
			Si & 0.1250 & 250 & 1.59  & 1294 & 0.849 & 161-179 & 188-205 \\ 
			P & 0.1250 & 200 & 2.02  & 1238 & 0.841  & 197-215 & 222-238 \\
			P & 0.1250 & 250 & 1.53  & 1520 & 0.922  & 180-199 & 207-224 \\
			P & 0.0625 & 200 & 2.29  & 1193 & 0.933 & 212-231 & 246-262 \\ 
			P & 0.0625 & 250 & 1.66  & 1512 & 0.955 & 196-216 & 227-244 \\
			Cl & 0.1250 & 200 & 1.31  & 1424 & 0.706  & 137-154 & 153-168 \\  
			Cl & 0.1250 & 250 & 1.08  & 1591 & 0.700  & 115-133 & 129-145 \\ \hline 
			S & - & 200 & 1.85 & 1358 & 0.868 & 195-214 & 222-238 \\ 
			S & - & 250 & 1.46 & 1571 & 0.875 & 175-195 & 197-219 \\  
		\end{tabular}
	\end{ruledtabular}
	\label{summary}
\end{table*}

As shown in previous studies \cite{PhysRevB.93.224513,doi:10.7566/JPSJ.87.124711}, the P doped $ \mathrm{H_3S} $ has the high superconductivity due to the enhanced density of states at the Fermi level of parent $ \mathrm{H_3S} $. We have also performed the simulations for elucidating the physical mechanism of this compound by using a supercell of 64 atoms ($ \mathrm{H_{48}S_{15}P} $). As is shown in Fig. \ref{p_cub}\hyperref[p_cub]{(a)} and Fig. S4, there are several flat bands along the $ \boldsymbol{k} $ path $ F\to Q\to Z $ at the Fermi surface, with the derivative $ \partial E_n/\partial k $ almost zero. This indicates that the doping of P alters the two van Hove singularities \cite{PhysRevB.93.104526,PhysRevB.93.224513} and can result in a peak of density of states right at the Fermi surface. Moreover, we found that the symmetry is preserved well after doping due to the existence of the near degeneracies in electronic band structures as shown in Fig. \ref{p_cub}\hyperref[p_cub]{(a)}, which can also be a critical factor to induce Van Hove singularities \cite{wang2021absence}. The high electronic density of states at the Fermi level could contribute to the large magnitude of phonon linewidth, as shown in Fig. \ref{p_cub}\hyperref[p_cub]{(b)} and contrasted with Fig. S5 \cite{SI}. We have also found the hybridization of $s$ orbitals of H with $p$ orbitals of S and P near the Fermi surface (Fig. \ref{p_cub}\hyperref[p_cub]{(c)}). The negligible curvature of the flat bands indicates well localization of corresponding $ s $ and $ p $ electrons, revealing the strong P-H and S-H chemical bondings. Fig. \ref{p_cub}\hyperref[p_cub]{(d)} shows the existence of [$ \mathrm{PH_{6}} $] and [$ \mathrm{SH_{6}} $] units, where the distances between P-H and S-H are compressed to 1.470 \r{A} and 1.474 \r{A}, compared with 1.493 \r{A} in $ \mathrm{H_3S} $ and 1.494 \r{A}-1.511 \r{A} for other S-H bonds in $ \mathrm{H_{48}S_{15}P} $. Finally, we found that the 6.25\% P doping could enhance the $ T_c $ of $ \mathrm{H_{3}S} $ around 20-30 K at 200 GPa and 250 GPa. 

To explore the superconductivity of the C-doped $ \mathrm{H_{3}S} $, we have computed the electron-phonon coupling strength of $ \mathrm{H_{32}S_{7}C} $, $ \mathrm{H_{36}S_{11}C} $ and $ \mathrm{H_{48}S_{15}C} $. For larger supercells corresponding to lower-proportion C-doped $ \mathrm{H_{3}S} $, the simulations could not be afforded due to computational demanding of these simulations as well as the limition of our computational power. Therefore, we attempt to estimate the superconductivity of low-proportion C-doped $ \mathrm{H_{3}S} $ without performing actual electron-phonon coupling simulations for a large supercell. Given that similar structures of the low-proportion C-doped $ \mathrm{H_{3}S} $ shares similar magnitude of average mass and electron-phonon interaction at the same pressure, we could thus estimate $ \lambda $ of the $ \mathrm{MgB_2} $ type superconductors by the Hopfield expression \cite{PhysRev.186.443,PhysRevLett.93.237002} 
\begin{equation}
	\lambda = \dfrac{N(0)I^2}{M\omega^2} \label{estimate1} 
\end{equation}
Then, the $ T_{c} $ can be estimated by \cite{PhysRevLett.93.237002}
\begin{equation}
	T_c = \omega\exp\left(-\dfrac{1}{\frac{\lambda}{1+\lambda}-\mu^{*}}\right) \label{estimate2}
\end{equation}
Further details of our estimation approach is also provided in the supplemental material \cite{SI}. As is shown in Table S2, values of $ T_{c} $ computed from our approach vary only about 5\% at maximum compared with that directly computed by QE, indicating the validity of our computational scheme. We thus investigated the superconductivity of the doping system of H$ {}_{3} $S up to 256 atoms as shown in Table S3 and Fig. \ref{c260}. The $ T_{c} $ of H$ {}_{3} $S$ {}_{1-x} $C$ {}_{x} $ with $ x $=0.1250-0.0156 could reach 148-192 K with $ \mu^{*} $=0.10 at 260 GPa. This suggests that the estimated superconductivity of C-doped $ \mathrm{H_{3}S} $ is much lower than room temperature as predicted in previous studies \cite{wang2021absence,DOGAN20211353851}. The reason for this discrepancy is possibly because different approches were employed. As for other predicted compounds, the superconductivity-related information is included in Table \ref{summary}.


\section{CONCLUSION}
In summary, we have computed the superconductivity of light-elements doped $ \mathrm{H_3S} $ using supercell approach within the framework of first-principle electronic structure. Our simulations indicate that the doping of Na can lower the dynamically stable pressure of $\mathrm{H_{3}S}$ while the doping of P can increase the density of states at the Fermi level as well as the superconductivity of $\mathrm{H_{3}S}$. Remarkably, we found that the existence of octahedra [$ \mathrm{PH_{6}} $] and [$ \mathrm{SH_{6}} $] units with squeezed P-H and S-H bonds in $ \mathrm{H_{48}S_{15}P} $ which are likely to be related to the high density of states at the Fermi level, with the higher $ T_{c} $ of 20-30 K compared with $ \mathrm{H_3S} $. Furthermore, we have proposed an estimation approach to reasonably estimate the superconductivity of the low proportion C-doping $ \mathrm{H_3S} $ without performing electron-phonon calculations on a quite large supercell. Our current work may inspire future work toward searching for high-temperature superconductivity in light-elements doping systems. 

\section*{ACKNOWLEDGMENTS}
This work was supported by the  Major  Program  of  the  National  Natural  Science  Foundation  of  China  (Grant  No.  52090024), National Natural Science Foundation of China (Grant No. 12074138, 11874175, and 11874176), Fundamental Research Funds for the Central Universities (Jilin University, JLU), Program for JLU Science and Technology Innovative Research Team (JLUSTIRT), and The Strategic Priority Research Program of Chinese Academy of Sciences (Grant No. XDB33000000). This work used computing facilities at the High-Performance Computing Centre of Jilin University.

\bibliographystyle{apsrev4-2}
\bibliography{apssamp}

\end{document}


\pagestyle{fancy}
\title{Supplemental Material for ``Superconductivity of Light-Elements Doped $ \mathbf{H_3S} $"}

\author{Hongyi Guan}
\affiliation{International Center for Computational Method $\&$ Software and State Key Laboratory of Superhard materials, College of Physics, Jilin University, Changchun 130012, China 
}
\author{Ying Sun}
\affiliation{International Center for Computational Method $\&$ Software and State Key Laboratory of Superhard materials, College of Physics, Jilin University, Changchun 130012, China 
}
\author{Hanyu Liu}
\email{hanyuliu@jlu.edu.cn}
\affiliation{International Center for Computational Method $\&$ Software and State Key Laboratory of Superhard materials, College of Physics, Jilin University, Changchun 130012, China 
}
\affiliation{Key Laboratory of Physics and Technology for Advanced Batteries (Ministry of Education), College of Physics, Jilin University, Changchun 130012, China}
\affiliation{International Center of Future Science, Jilin University, Changchun 130012, China}
\maketitle

\section*{Supplemental Analysis}
\subsection{Mcmillan-Allen-Dynes Equation}
The Mcmillan-Allen-Dynes equation can be expressed as \cite{PhysRevB.12.905}
\begin{equation}
T_c = f_{1}f_{2}\dfrac{\omega_{\mathrm{log}}}{1.20}\exp\left[-\dfrac{1.04(1+\lambda)}{\lambda-\mu^{*}(1+0.62\lambda)}\right] \label{tc}
\end{equation}
Where 
\begin{gather}
\lambda = \int \dfrac{2\alpha^2F(\omega)}{\omega}\mathrm{d} \omega \label{lambda} \\
\omega_{\mathrm{log}} = \mathrm{exp}\left[\dfrac{2}{\lambda}\int \dfrac{\alpha^2F(\omega)}{\omega}\log\omega \mathrm{d} \omega\right] \label{wlog}
\end{gather} 
$\lambda$ and $\omega_{\mathrm{log}}$ are isotropic electron-phonon coupling parameter and logarithmically averaged phonon frequency, respectively.  $ \lambda $ depends on a critical quantity 
\begin{equation}
\alpha^2F(\omega) = \dfrac{1}{N(0)}\sum\limits_{i,j,\nu}\iint \dfrac{\mathrm{d} \boldsymbol{k}}{\Omega}\dfrac{\mathrm{d} \boldsymbol{q}}{\Omega}|g_{ij\nu}(\boldsymbol{k},\boldsymbol{q})|^2\delta (\omega-\omega_{\boldsymbol{q}\nu})\delta (\varepsilon_{i\boldsymbol{k}}-\varepsilon_F)\delta (\varepsilon_{j\boldsymbol{k}+\boldsymbol{q}}-\varepsilon_F) \label{a2f}
\end{equation}
Where $ \Omega $ is the volume of the first Brillouin zone, and 
\begin{equation}
g_{ij\nu}(\boldsymbol{k},\boldsymbol{q}) = \sqrt{\dfrac{\hbar}{2M\omega_{\boldsymbol{q}\nu}}}\bra{i,\boldsymbol{k}}\partial_{\boldsymbol{q}\nu}V^{\text{scf}}\ket{j,\boldsymbol{k}+\boldsymbol{q}}
\end{equation}
$g_{ij\nu}(\boldsymbol{k},\boldsymbol{q})$ is the electron phonon matrix element\cite{PhysRevB.12.905,PhysRevB.9.4733}.
To analyze the impact of each quantity on $ T_c $, we first rewrite Eq. \ref{tc} in a simplified form notice correction factor $ f=f_{1}f_{2}\sim 1 $
\begin{equation}
T_c = \dfrac{\omega_{\mathrm{log}}}{1.20}\exp\left[-\dfrac{1.04(1+\lambda)}{\lambda-\mu^{*}(1+0.62\lambda)}\right]
\end{equation}
Since $ \omega_{\mathrm{log}} $ is nearly irrelevant to the magnitude of $ \alpha^2F $, thus the change of magnitude of $ \alpha^2F $ only affect $ \lambda $. Thus setting the screened Coulomb parameter $ \mu^{*} $ fixed, we have 
\begin{equation}
\dfrac{\partial T_c}{\partial \lambda} = \dfrac{\omega_{\mathrm{log}}}{1.20}\exp\left[-\dfrac{1.04(1+\lambda)}{\lambda-\mu^{*}(1+0.62\lambda)}\right]\dfrac{1.04+0.3952\mu^{*}}{[\lambda-\mu^{*}(1+0.62\lambda)]^2} > 0
\end{equation}
Furthermore, the larger $ \lambda $ enables $ f $ larger as well. Thus the increase of magnitude of $ \alpha^2F $ will enhance $ T_c $. Note that 
\begin{equation}
\iint \dfrac{\mathrm{d} \boldsymbol{k}}{\Omega}\dfrac{\mathrm{d} \boldsymbol{q}}{\Omega}\delta (\varepsilon_{i\boldsymbol{k}}-\varepsilon_F)\delta (\varepsilon_{j\boldsymbol{k}+\boldsymbol{q}}-\varepsilon_F) \sim N(0)^2 \quad \int \delta (\omega-\omega_{\boldsymbol{q}\nu})\mathrm{d} \omega \sim N_{\mathrm{ph}}(\omega)
\end{equation}
Thus regardless of anisotropy, $ \lambda \sim |g|^2N(0)N_{\mathrm{ph}}(\omega) $. By further inspecting Eq. \ref{a2f}, we found that if at specific area (in the direct product of two reciprocal spaces $ V_{\boldsymbol{k}}\otimes V_{\boldsymbol{q}} $) the electron density of states, phonon density of states and electron phonon coupling are simultaneously large, the $ \lambda $ can be greatly enhanced while keeping $ \omega_{\mathrm{log}} $ at large values. If we only focus on $ \boldsymbol{q} $, we can integrate by $ \boldsymbol{k} $ and define 
\begin{equation}
\gamma_{\boldsymbol{q}\nu} = 2\pi \omega_{\boldsymbol{q}\nu}\sum\limits_{i,j}\int \dfrac{\mathrm{d} \boldsymbol{k}}{\Omega}|g_{ij\nu}(\boldsymbol{k},\boldsymbol{q})|^2\delta (\varepsilon_{i\boldsymbol{k}}-\varepsilon_F)\delta (\varepsilon_{j\boldsymbol{k}+\boldsymbol{q}}-\varepsilon_F)
\end{equation}
We can obtain below expression
\begin{equation}
\alpha^2F(\omega) = \dfrac{1}{2\pi N(0)}\sum _{\nu}\int\dfrac{\mathrm{d} \boldsymbol{q}}{\Omega}\delta (\omega - \omega_{\boldsymbol{q}\nu})\dfrac{\gamma_{\boldsymbol{q}\nu}}{\hbar \omega_{\boldsymbol{q}\nu}}
\end{equation}
We found that $ \lambda \propto N_{ph}\gamma $.
\subsection{Estimation Approach of C-Doped $ \mathbf{H_{3}S} $ and Its Validity}
For large supercells (H$ {}_{3} $S$ {}_{1-x} $C$ {}_{x} $), we computed the density of states at Fermi surface $ N(0)_{x} $, then given that $ T_{c} $ have a roughly ``$ \omega^2 $ free" expression\cite{PhysRevB.12.905,PhysRevB.93.104526} 
\begin{equation}
	T_c = 0.18\sqrt{\lambda\omega^2} = 0.18\sqrt{N(0)I^2/M}
\end{equation}
in Allen-Dynes limit and we assume similar structures share similar scattering matrix element $I^2$ and similar average inonic mass, we can define a pseudo ``electron-phonon coupling constant" $ \lambda_{x} $ by the $ \lambda $ from a known structure: $ \lambda_{x} \equiv \lambda N(0)_{x}/N(0) $. Here $ N(0) $ is the density of states of known structure at the Fermi surface. For the reason of self-consistency, we use the density of states calculated without considering electron-phonon interpolation to estimate all the $ N(0) $ and $ N(0)_{x} $. Then with $ T_{c} $ of the known structure, the transition temperature $ T_{cx} $ is estimated to be
\begin{equation}
	T_{cx}=T_{c}\exp\left(-\dfrac{1}{\frac{\lambda_{x}}{1+\lambda_{x}}-\mu^{*}}\right)/\exp\left(-\dfrac{1}{\frac{\lambda}{1+\lambda}-\mu^{*}}\right)
\end{equation} 
We estimate the $ T_{cx} $ of H$ {}_{3} $S$ {}_{1-x} $C$ {}_{x} $ by both the 64 atoms supercell $ \mathrm{H_{3}S_{0.9375}C_{0.0625}} $ and $ \mathrm{H_{3}S} $, to obtain the values $ T_{cx,0.0625} $ and $ T_{cx,0} $ resprectively and estimate our final result as 
\begin{equation}
	T_{cx} = \dfrac{x}{0.0625}T_{cx,0.0625}+\left(1- \dfrac{x}{0.0625}\right)T_{cx,0}
\end{equation}

To illustrate the validity of this approach, we estimate the $ T_c $ of $ \mathrm{H_{3}S_{0.9375}X_{0.0625}} $ by $ \mathrm{H_{3}S_{0.875}X_{0.125}} $ and $ \mathrm{H_{3}S} $, in this scenario, $ 	T_{cx} = \dfrac{x}{0.125}T_{cx,0.125}+\left(1- \dfrac{x}{0.125}\right)T_{cx,0} $. All the results from the estimated approach and directly computed by Quantum Espresso are shown in Table \ref{validity}, where their difference is only around 5\% at maximum.
\section*{Supplemental Figures}
\begin{figure}[H]
\centering
	\includegraphics[width=0.6\linewidth]{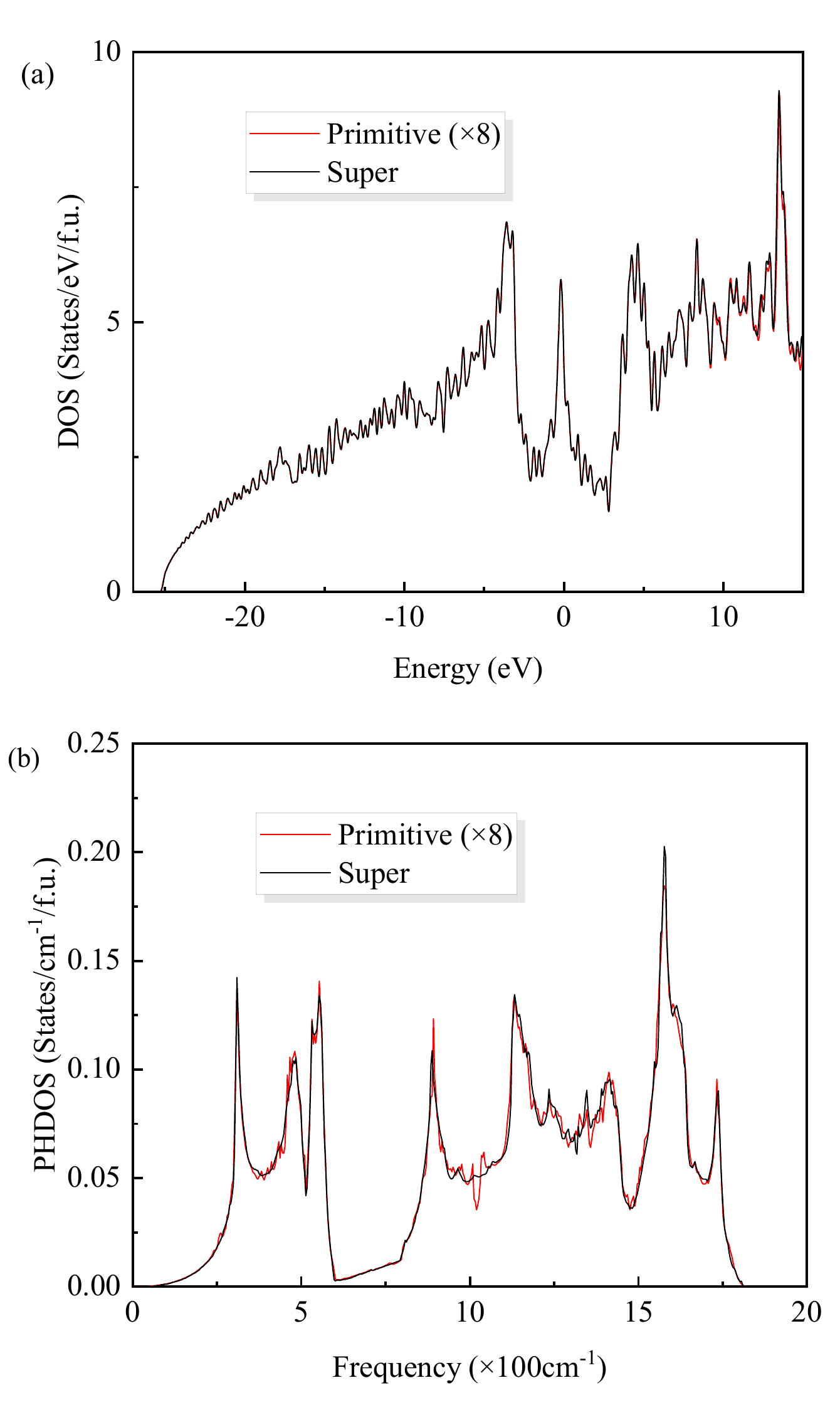}
\caption{Comparison of (a) electronic density of states and (b) phonon density of states between primitive cell and a supercell of 32 atoms at 200 GPa.}
\end{figure}
\newpage
\begin{figure}[H]
	\centering
	\includegraphics[width=0.6\linewidth]{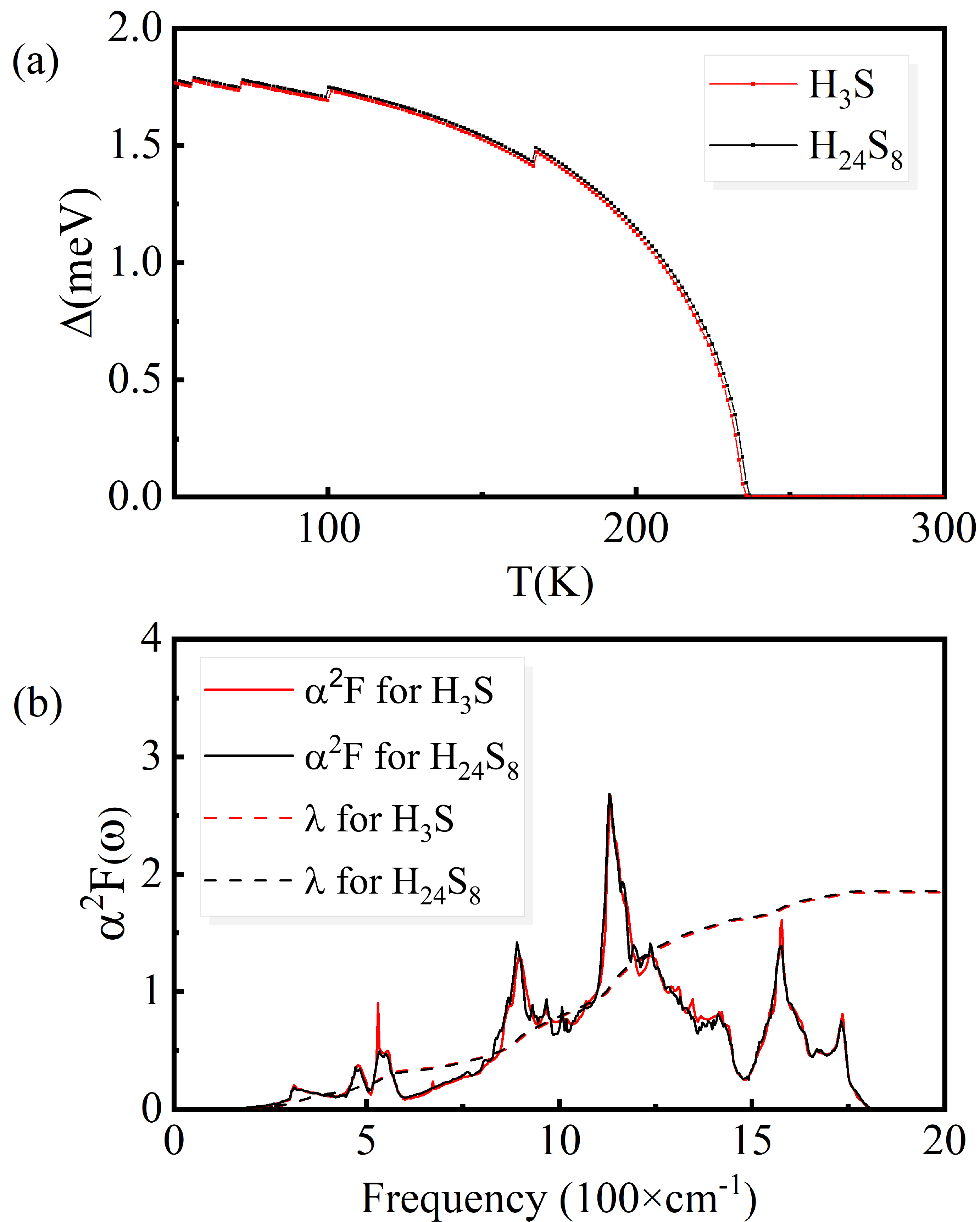}
	\caption{Comparison of (a) superconducting gap ($ \mu^{*} = 0.10 $) and (b) Eliashberg spectral function of H$ {}_{3} $S computed by primitive cell and a supercell of $ 2\times 2\times 2 $ primitive cell  at 200 GPa.}
\end{figure}

\begin{figure}[H]
	\centering
	\includegraphics[width=\linewidth]{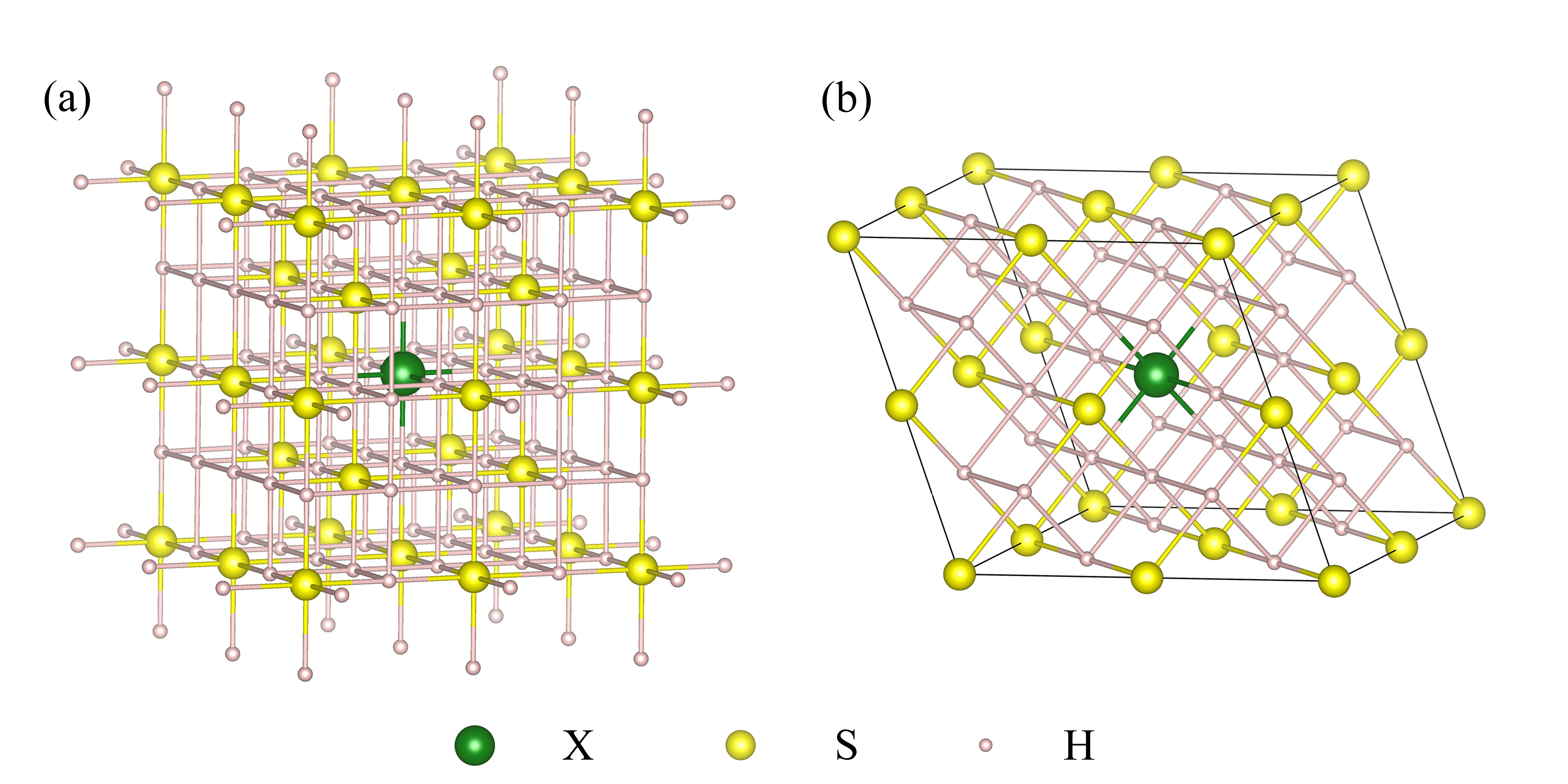}
	\caption{(a) Structure of $ \mathrm{H_{48}S_{15}X} $ (b) Structure of $ \mathrm{H_{24}S_7X} $. X denotes the doped elements.}
	\label{sp222}
\end{figure}

\begin{figure}[H]
\centering
	\includegraphics[width=.7\linewidth]{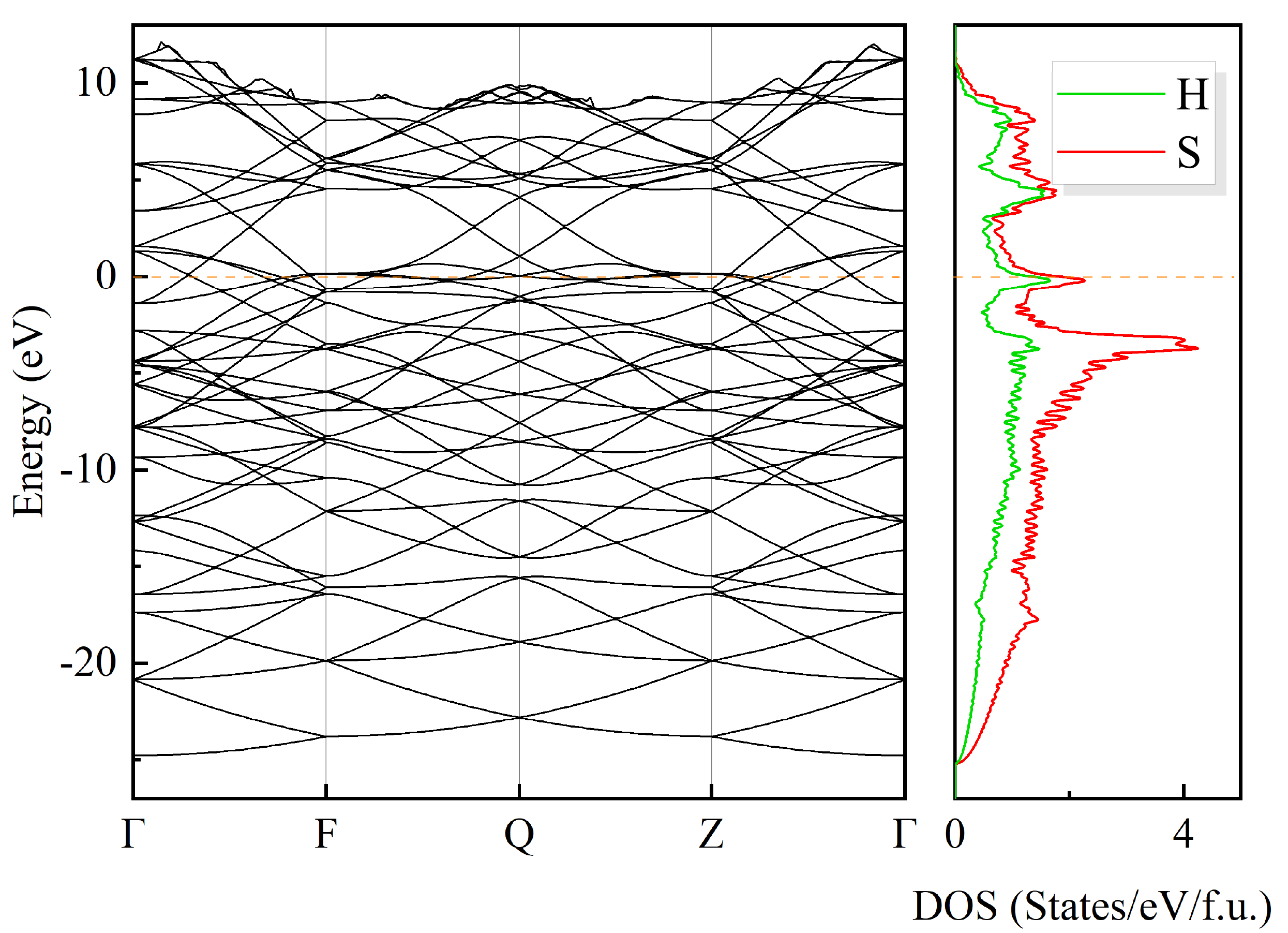}
	\caption{The electronic band structure and projected density of states of $ \mathrm{H_{48}S_{16}} $ at 200 GPa.}
\end{figure}

\begin{figure}[H]
\centering
	\includegraphics[width=.7\linewidth]{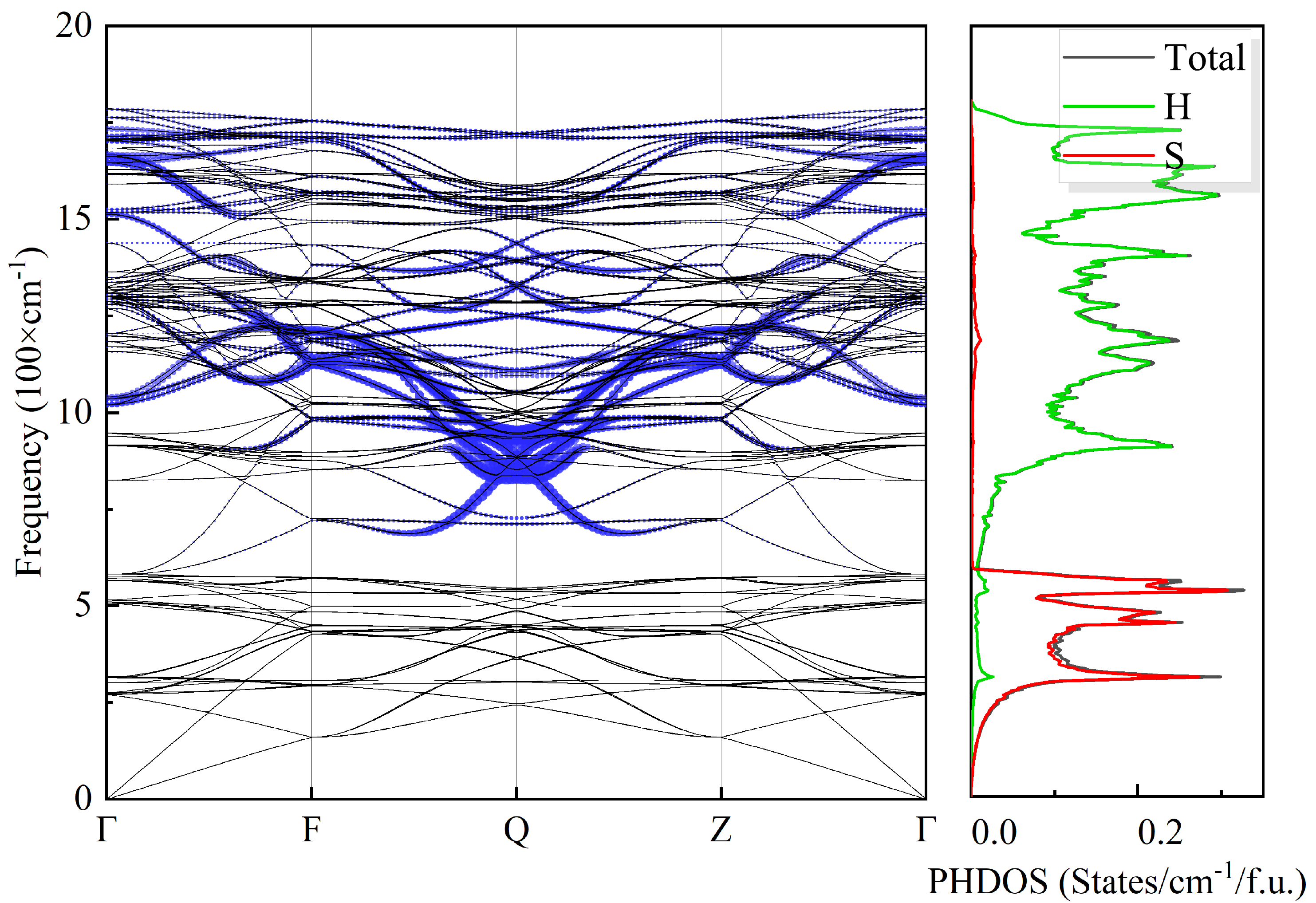}
	\caption{The phonon dispersion relationship with phonon linewidth (left panel) and phonon density of states (right panel) of $ \mathrm{H_{48}S_{16}} $ at 200 GPa. The radii of the blue circles indicate the magnitude of the phonon linewidth.}
\end{figure}

\section*{Supplemental Tables}
\begin{table}[H]
\centering
 \caption{Comparison of superconductivity related parameters of $ \mathrm{H_3S} $ and $ \mathrm{H_{24}S_{8}} $ with the screened Coulomb parameter $ \mu^{*} $ of 0.10-0.13.}
\begin{ruledtabular}
\begin{tabular}{ccccccc}
        Formula & Pressure (GPa) & $\lambda$ & $\omega_{\mathrm{log}}$ (K) & \makecell[c]{$N(0)$ \\ (States/Ry/atom)} & \makecell[c]{$T_c$ by \\ MAD equation (K)} & \makecell[c]{$T_c$ by \\ ME eqaution (K)} \\ \hline
          $ \mathrm{H_3S} $ & 200 & 1.85 & 1358 & 0.868 & 195-214 & 222-238 \\
          $ \mathrm{H_{24}S_{8}} $ & 200 & 1.86 & 1358 & 0.867 & 196-215 & 223-239 \\ 
    \end{tabular}
\end{ruledtabular}
    \label{compare}
\end{table}

\begin{table}[H]
	\centering
	\caption{Density of states at the Fermi surface without considering electron-phonon interpolation (the subscripts denote doping ratio) and the comparison of the superconducting transition temperature of $ \mathrm{H_{3}S_{0.9375}X_{0.0625}} $ derived from the estimation approach and computed by Quantum Espresso with $ \mu^{*} $ of 0.10-0.13.}
	\begin{ruledtabular}
		\begin{tabular}{ccccccc}
			Dopant X & Pressure (GPa) &  \makecell[c]{$N(0)_{0.0625}$ \\ (States/Ry/atom)} & \makecell[c]{$N(0)_{0}$ \\ (States/Ry/atom)} & \makecell[c]{$N(0)_{0.125}$ \\ (States/Ry/atom)} &
			\makecell[c]{Estimated \\ $ T_{c} $ (K)} & 
			\makecell[c]{Computed \\ $ T_{c} $ (K)}
			\\ \hline
            C & 260 & 0.867 & 0.941 & 0.744 & 157-175 & 154-170 \\ 
            P & 250 & 0.947 & 0.939 & 0.897 & 184-203 & 187-207 \\ 
            P & 200 & 0.950 & 0.919 & 0.913 & 202-220 & 212-231 \\
		\end{tabular}
	\end{ruledtabular}
	\label{validity}
\end{table}

\begin{table}[H]
	\centering
	\caption{Density of states at the Fermi surface without considering electron-phonon interpolation and superconducting transition temperature of H$ {}_{3} $S$ {}_{1-x} $C$ {}_{x} $ at 260 GPa with $ \mu^{*} $ of 0.10-0.13. The $ x $ marked by ``*" means the values of $ T_{cx} $ are estimated.}
	\begin{minipage}[c]{0.6\linewidth}
		\begin{ruledtabular}
			\begin{tabular*}{0.5\linewidth}{cccc}
				Doping ratio $ x $ & $ \lambda $ & \makecell[c]{$N(0)_{x}$ \\ (States/Ry/atom)} & $ T_{c} $ (K) \\ \hline
				0.1250  & 1.34 & 0.744 & 132-148 \\ 
				0.0833  & 1.52 & 0.866 & 154-173 \\ 
				0.0625  & 1.80 & 0.867 & 154-170 \\ 
				0.0556$ ^{*} $ & - & 0.892 & 158-175 \\ 
				0.0417$ ^{*} $ & - & 0.905 & 161-179 \\ 
				0.0370$ ^{*} $ & - & 0.895 & 160-177 \\ 
				0.0313$ ^{*} $ & - & 0.898 & 161-179 \\ 
				0.0278$ ^{*} $ & - & 0.902 & 162-180 \\ 
				0.0208$ ^{*} $ & - & 0.920 & 165-185 \\ 
				0.0156$ ^{*} $ & - & 0.938 & 170-188 \\ 
				0.0000  & 1.42 & 0.941 & 172-192 \\ 
			\end{tabular*}
		\end{ruledtabular}
	\end{minipage}
	\label{c260}
\end{table}

\bibliography{Supplemental_Material.bib}